\documentclass[usenatbib]{mnras}
\usepackage{newtxtext,newtxmath}
\usepackage{amsmath}
\usepackage{ctable}
\usepackage{verbatim}
\usepackage{url}
\usepackage{fixltx2e} 
\usepackage{graphicx}
\usepackage{subcaption}
\usepackage{bm}
\usepackage{stfloats}
\usepackage{ulem}
\newcommand{\dataavailability}[1]{\begin{small}\section*{Data
Availability}\end{small}{\noindent #1}\vspace{5pt}}

\newcommand{\myemail}{suoqing@shao.ac.cn}

\title[Tayler Instability in Rotating Stellar Interiors]{Magnetohydrodynamic Simulations of the Tayler Instability in Rotating Stellar Interiors}

\author[Ji, Fuller \& Lecoanet]
  {Suoqing Ji,$^{1,2}$\thanks{\myemail} Jim Fuller$^2$ and Daniel
  Lecoanet$^{3,4}$ \\
  $^1$Astrophysics Division \& Key Laboratory for Research in Galaxies and
  Cosmology, Shanghai Astronomical Observatory, Chinese Academy of Sciences,\\
  Shanghai 200030, China\\
  $^2$TAPIR, Walter Burke Institute for Theoretical Physics, California
  Institute of Technology, Pasadena, CA 91125, USA\\
  $^3$Department of Engineering Sciences and Applied Mathematics, Northwestern
  University, Evanston IL 60208, USA \\
  $^4$CIERA, Northwestern University, Evanston IL 60201, USA }

\date{}
\begin{document}
\maketitle

\begin{abstract}

The Tayler instability is an important but poorly studied magnetohydrodynamic
instability that likely operates in stellar interiors. The nonlinear saturation
of the Tayler instability is poorly understood and has crucial consequences for
dynamo action and angular momentum transport in radiative regions of stars. We
perform three-dimensional MHD simulations of the Tayler instability in a
cylindrical geometry, including strong buoyancy and Coriolis forces as
appropriate for its operation in realistic rotating stars. The linear growth of
the instability is characterized by a predominantly $m=1$ oscillation with
growth rates roughly following analytical expectations. The non-linear
saturation of the instability appears to be caused by secondary shear
instabilities and is also accompanied by a morphological change in the flow. We
argue, however, that non-linear saturation likely occurs via other mechanisms in
real stars where the separation of scales is larger than those reached by our
simulations. We also observe dynamo action via the amplification of the
axisymmetric poloidal magnetic field, suggesting that Tayler instability could
be important for magnetic field generation and angular momentum transport in the
radiative regions of evolving stars.

\end{abstract}


\section{Introduction}
\label{sec:intro}

The interplay between rotation and magnetism is crucial for understanding the
evolution of stars and the compact objects they produce. Differential rotation
generated by contracting stellar cores may source various magnetohydrodynamic
(MHD) instabilities that can amplify magnetic fields and/or transport AM
outwards to slow the rotation of the stellar core. However, the instabilities at
work and their saturation mechanisms remain poorly understood.

Asteroseismic observations have helped by providing internal rotation rate
measurements for stars on the main sequence
\citep{kurtz:14,saio:15,benomar:15,vanreeth:18}, sub-giant/red giant branch
(RGB) \citep{beck:12,mosser:12,deheuvels:14,triana:17,gehan:18}, red clump
\citep{mosser:12,deheuvels:15}, and finally in WD remnants \citep{hermes:17}.
The conclusion drawn from these measurements is unambiguous: core rotation rates
are relatively slow, and the vast majority of AM is extracted from stellar cores
as they evolve. The spin rates of red giant cores and WDs are slower than
theoretically predicted by nearly all hydrodynamic AM transport mechanisms
\citep{fullerwave:14,belkacem:15,spada:16,eggenberger:17,ouazzani:19}.

The non-axisymmetric MHD Tayler instability
\citep{tayler:73,spruit:99,goldstein:19} is likely the most important MHD
instability in radiative regions of stars. Tayler instability is a kink-type
instability of toroidal (azimuthal) fields that have been created by winding up
a radial seed field through differential rotation. Above a critical field
strength, field loops slip sideways relative to one another with a predominantly
non-axisymmetric $m=1$ wavenumber. While the dynamics of the linear instability
are well understood, the nonlinear (and likely turbulent) saturation of the
instability, and the resulting AM transport are poorly understood and
controversial \citep[e.g.,][]{braithwaite:06,zahn:07}.

The Tayler-Spruit (TS) dynamo \citep{spruit:02} is one possible saturation
mechanism of the Tayler instability. In this theory, toroidal magnetic field
energy is turbulently dissipated by the fluid motions, and the instability
saturates when the turbulent dissipation rate is equal to the energy input via
winding of the radial field. In the presence of a composition gradient, the
resulting torque density due to Maxwell stresses is
\begin{align}
T_{\rm TS} \sim \rho r^2 \Omega^2 q^3 \left(\frac{\Omega}{N_{\rm eff}}\right)^4 \, ,
\label{eq:ts}
\end{align}
where $q = d \ln \Omega/d \ln r$ is the dimensionless radial shear, and $N_{\rm
eff}$ is the effective stratification, which is usually nearly equal to the
compositional stratification $N_\mu$ in post-MS stars. The TS dynamo has been
implemented into many stellar evolution codes, but it predicts much faster core
rotation than observed in post-MS stars \citep{cantiello:14}.

However, \citet{fuller:19} argued that \citet{spruit:02} overestimated the
energy damping rate of the instability, because only magnetic energy in the
disordered (perturbed) field can be turbulently damped. By calculating an energy
damping rate due to weak magnetic turbulence, \citet{fuller:19} argued the
instability can grow to larger amplitudes, producing a larger Maxwell stress in
its saturated state. \citet{fuller:19} find the Tayler torques are 
\begin{align}
T_{\rm TSF} = \alpha^3 \rho r^2 \Omega^2 q \left(\frac{\Omega}{N_{\rm eff}}\right)^2 \, ,
\label{eq:tsf}
\end{align}
where $\alpha$ is a saturation parameter of order unity. The different scaling
is very important because $\Omega/N_{\rm eff} \sim 10^{-4}$ in RGB stars, so the
prescription of \citet{fuller:19} allows for significantly more AM transport.

Because the saturation of the Tayler instability is a complex and nonlinear
process, it is important to examine this process via numerical simulations. The
Tayler instability has been seen in a few simulations, but only in limited
configurations not including both rotation and realistic stratification.
\citet{weber:15} and \citet{gellert:08} used $N=0$ (i.e., no stratification) and
\citet{guerrero:19} used $\Omega=0$ (i.e., no rotation). The first simulation of
the Tayler instability with shear and buoyancy \citep{braithwaite:06} was
compressible, limiting the dynamic range and time scale over which simulations
could be performed. Those simulations used $\Omega/N = 1$, in stark contrast to
the values of $\Omega/N \sim 10^{-4} \ll 1$ expected in real stars. The analytic
predictions of \citet{spruit:02} and \citet{fuller:19} also assume $\Omega \ll
N$, so it is important to simulate that parameter regime. Recently,
\citet{petitdemange:23} presented a suite of simulations of Tayler instability,
finding apparent agreement with the prediction of Eq.~\eqref{eq:ts}, which we
discuss further in Section \ref{sec:diss}.

In this work, we perform three-dimensional MHD simulations of the Tayler
instability, including both stratification and rotation. We also vary
dimensionless parameters over a small range in an attempt to determine scaling
relations and extrapolate the nature of the saturated state to parameters
characteristic of real stars. Our paper is organized as follows. The numerical
methods and selected parameters are described in \S\ref{sec:method}. In
\S\ref{sec:results}, we discuss the results regarding the linear and non-linear
evolution of the Tayler instability. We finally conclude in \S\ref{sec:diss}.

\section{Methods \& Simulation Setup}
\label{sec:method}

\subsection{Simulation code}

We use the spectral MHD code {\small
Dedalus}\footnote{\url{http://dedalus-project.org}} \citep{burns:20} for our
simulations. We use version 2.2006 of the code with commit hash {\tt 9bf7eb1}.
Because of its spectral nature, {\small Dedalus} can achieve comparable accuracy
with relatively lower resolutions compared with extremely high-resolution
simulations using finite-volume codes, and it parallelizes efficiently using
MPI. This feature is particularly useful for our 3D simulations, since only in
three dimensions can the Tayler instability develop. {\small Dedalus} has
already demonstrated its ability to handle different types of MHD problems
including effects of stratification and nearly incompressible dynamics including
convection, waves, and magnetic fields
\citep[e.g.,][]{lecoanet:15,lecoanet:17,couston:18}.

\subsection{Initial conditions}

For convenience, we non-dimensionalize the initial conditions by setting the
characteristic scales (width $L_\mathrm{box}$, averaged gas density $\bar{\rho}$
and gravity $g$) to unity $1$. To mimic a latitudinal band of a star, our
simulations are performed in 3D cylindrical coordinates $(r, z, \phi)$ with the
domain size of
\begin{gather}
    R_\mathrm{in} \leq r \leq R_\mathrm{out} \\
    -Z \leq z \leq Z \\
    0 \leq \phi \leq 2\pi,
\end{gather}
where $R_\mathrm{in} = L_\mathrm{box}$, $R_\mathrm{out} = 2 L_\mathrm{box}$ and
$Z = L_\mathrm{box} / 2$ \footnote{The simulation domain has an aspect ratio of
$1$ in $r$-$z$ plane. We choose this aspect ratio for convenience, and the wave
numbers along $r$ and $z$ directions can also be sufficiently resolved with this
aspect ratio, given that $k_z \sim 2 \pi N / \omega_\mathrm{A} r$ (where the
Alfv\'en frequency $\omega_\mathrm{A} \equiv B /\sqrt{\bar{\rho}r}$) and $k_r
\sim 2 \pi/L_\mathrm{box}$ are at similar orders of magnitude for parameters
used in our simulations (see the following \S\ref{sec:sim_params}).}. We set up
initial conditions as a magnetized, gravitationally stratified medium with
density gradient $d\rho_0/dz$ and gravitational acceleration $\bm{g}$ along the
$z$-axis:
\begin{gather}
    \rho_0(z) = \bar{\rho} + \frac{d\rho_0}{dz} z \label{eq:rho_z} \\
    \bm{g} = - g \hat{\bm{e}}_z \, .
\end{gather}
Here, $\rho_0$ denotes the density of the unperturbed state which is a function
of the scale height $z$. The averaged density $\bar{\rho}$, and the density
gradient $d\rho_0 / dz$ are constants.

We initialize the simulation with a toroidal magnetic field $\bm{B}$ along
$\phi$-direction, with a power-law profile in radius:
\begin{equation}
    \bm{B}(r) = B_0 \left(\frac{r}{R_\mathrm{in}}\right)^p \hat{\bm{e}}_\phi \, ,
    \label{eq:B0}
\end{equation}
where $B_0$ and $p$ are constant. Here we use $p=2$ for the initial magnetic
field profile which is expected to be Tayler unstable \citep{tayler:73}. The
system is initially in magnetostatic equilibrium with the unperturbed pressure
$p_0$ satisfying:
\begin{gather}
    \frac{\partial (p_0 + B^2/2)}{\partial r} = \left|(\bm{B}\cdot \nabla) \bm{B} \right|_r \\
    \frac{\partial p_0}{\partial z} = -\rho_0 g \\
    \frac{\partial p_0}{\partial \phi} = 0,
\end{gather}
such that pressure gradients are balanced by magnetic forces and gravity
respectively in the $r$ and $z$ directions.

\subsection{Initial perturbations}

Since the Tayler instability is a non-axisymmetric instability, initially
non-axisymmetric perturbations are needed, otherwise, perfect axisymmetry will be
maintained throughout the simulations. To maintain $\nabla \cdot \bm{B} = 0$, we
effectively evolve the magnetic vector potential $\bm{A}$ in our equation sets
with $\bm{B} \equiv \nabla \times \bm{A}$. We initialize white-noise
perturbations to the magnetic fields by setting the magnetic potential vector
$\bm{A}$ as:
\begin{align}
    \bm{A} = 10^{-10} B_0 W[0,1] r \hat{\bm{e}}_r - \frac{B_0 r^{p+1}}{(p+1)R_\mathrm{in}^p} \hat{\bm{e}}_z, \label{eq:mag_pot}
\end{align}
where $W[0,1]$ is a random number uniformly distributed between $0$ and $1$. By
taking $\bm{B} = \nabla\times \bm{A}$, we obtain divergence-free magnetic fields
with $B_\phi$ in desired form in Eq. \eqref{eq:B0} and white noise perturbations
with a magnitude of $\sim \! 10^{-10} B_0$ in $B_z$ and $B_\phi$. Because the
white noise distribution does not introduce any characteristic length scales,
these initial conditions do not add to the initial magnitude of any particular
modes.

\subsection{Governing equations}

We express fluid quantities as the sum of the unperturbed fields (denoted by the
subscript $0$) and the variations (denoted by the prime symbol), e.g., $\rho =
\rho_0 + \rho'$, $p = p_0 + p'$, etc., and solve the following fundamental
governing equations of incompressible magnetohydrodynamics:
\begin{gather}
    \frac{D\rho'}{D t}
        =  \frac{\bar{\rho}}{g} N^2 u_z 
            + \kappa \nabla^2\rho' \label{eq:rho_t} \\
    \frac{D\bm{u}}{D t}
        = -\nabla\left(\frac{p'}{\bar{\rho}}\right) 
        + \frac{\rho'}{\bar{\rho}} \bm{g}
        + \frac{(\nabla \times \bm{B})\times\bm{B}}{\bar{\rho}}
        - 2\bm{\Omega_0}\times\bm{u}
        + \nu \nabla^2\bm{u} \label{eq:momentum} \\
    \nabla \cdot \bm{u} = 0 \\
    \frac{\partial{\bm{B}}}{\partial t} 
        = \nabla\times\left(\bm{u}\times\bm{B}\right) + \eta \nabla^2\bm{B} \label{eq:induction} \\
    \nabla \cdot \bm{B} = 0, \label{eq:divB}
\end{gather}
where $D/Dt$ denotes $\partial /\partial t + \bm{u} \cdot \nabla$, $\bm{u}$ is
the fluid velocity, and $N$ is the Brunt-V\"{a}is\"{a}l\"{a} frequency defined
as:
\begin{gather}
    N^2 \equiv -\frac{\partial\rho_0}{\partial z}\frac{g}{\bar{\rho}}.
\end{gather}

We use a Boussinesq approximation that $|\rho - \bar{\rho}|\ll |\bar{\rho}|$,
which is acceptable because of the incompressible nature of the Tayler
instability and its short radial length scale. Vertical stratification appears
through the buoyancy term, which appears in spite of the Boussinesq
approximation. This mimics a simulation of a star over a radial length scale
much less than the density scale height. We transform our simulations into the
rotating frame by adding the Coriolis term $2\bm{\Omega_0}\times\bm{u}$, with
bulk angular velocity $\bm{\Omega}_0 = \Omega_0 \hat{\bm{e}}_z$. We include
explicit diffusivity, viscosity and magnetic resistivity as $\kappa$, $\nu$ and
$\eta$ respectively, where the diffusivity mimics the compositional diffusivity
in a real star. Temperature perturbations are not included because the
instability operates in an isothermal regime in post-MS stars. Since the
magnetic diffusivity is usually larger than microscopic viscosity in real stars,
we adopt a relatively large magnetic diffusivity $\eta$ with $\eta > \kappa \sim
\nu$.

\subsection{Boundary conditions}

We apply periodic boundary conditions along the $z$ and $\phi$-directions. Note
that although a density profile in the $z$ direction is implied as described by
Eq. \eqref{eq:rho_z}, what actually solved in the governing equations
\eqref{eq:rho_t} -- \eqref{eq:divB} are variations of fluid quantities (e.g.,
$\rho'$, $p'$, $\bm{u}$, $\bm{B}$, etc.), therefore periodic boundary conditions
can be applied to the $z$-axis. We apply $\rho'=0$, $p'=0$ and $\bm{u}=0$ at the
inner and outer boundaries. We apply the electric scalar potential $\phi_E = 0$
and the magnetic potential $A_\phi = 0$ and on both inner and outer boundaries,
and $A_z (r=R_\mathrm{in}) = -B_0 (p+1)^{-1} R_\mathrm{in} $ and $A_z
(r=R_\mathrm{out}) = -B_0 (p+1)^{-1} R_\mathrm{out}^{p+1} R_\mathrm{in}^{-p}$ at
the inner and outer boundaries respectively, in order to maintain continuity in
the magnetic potential described by Eq. \eqref{eq:mag_pot}. These boundary conditions are consistent with the initial field profile but allow the magnetic field to evolve. The boundary conditions enforce the toroidal magnetic flux to be conserved, but the magnetic energy can decrease (although it cannot go to zero).

\subsection{Simulation parameters}
\label{sec:sim_params}

\begin{table*}
    \renewcommand{\arraystretch}{1.5}

    \begin{center}
        \begin{tabular}{ccccccc}
          \toprule[1pt]\midrule[0.4pt] & Resolution & Averaged density
             $\bar{\rho}$ & Gravity $g$ & Diffusivity $\kappa$ & Viscosity $\nu$
             & Magnetic resistivity $\eta$ \\
             \cmidrule[0.4pt](lr){2-7} Value & $512\times 512 \times 64$ & $1$ &
        $1$ & $10^{-5}$ & $10^{-5}$ & $2\times 10^{-5}$ \\
             \cmidrule[0.4pt](lr){2-2} \cmidrule[0.4pt](lr){5-7} Note & in
        $r\times z \times \phi$ & & & \multicolumn{3}{c}{
                  \begin{tabular}{@{}c} corresponds to the Prandtl number of
                    $\mathrm{Pr} \equiv \frac{\nu}{\kappa} = 1$ \\ 
                    and the magnetic Prandtl number of $\mathrm{Pr_m} \equiv
                    \frac{\nu}{\eta} = 0.5$
                  \end{tabular}
            
                 }   \\ 
        \bottomrule[1pt]
        \end{tabular}
    \end{center}

    \vspace{0.2cm}

    \begin{center}
        \begin{tabular}{ccccccccccccc}
          \toprule[1pt]\midrule[0.4pt] & \multicolumn{5}{c}{Angular frequency
             $\Omega_0$} & \multicolumn{3}{c}{Alfv\'en frequency
             $\omega_\mathrm{A}$} & \multicolumn{4}{c}{Brunt-V\"{a}is\"{a}l\"{a}
             frequency $N$} \\
             \cmidrule[0.4pt](lr){2-6} \cmidrule[0.4pt](lr){7-9}
          \cmidrule[0.4pt](lr){10-13} Value & $0.3$ & $0.4$ & $0.5$ & $0.6$ &
          $0.7$ 
            & $0.2$ & $0.25$ & $0.3$ 
            & $0.9$ & $1$ & $1.3$ & $1.5$ \\ 
          Name & \tt Om.3 & \tt Om.4 & \tt Om.5 & \tt Om.6 & \tt Om.7
            & \tt OmA.2 & \tt OmA.25 & \tt OmA.3 
            & \tt N.9 & \tt N1 & \tt N1.3 & \tt N1.5 \\
            \cmidrule[0.4pt](lr){2-6} \cmidrule[0.4pt](lr){7-9}
        \cmidrule[0.4pt](lr){10-13} Note & \multicolumn{5}{c}{angular velocity
        $\bm{\Omega} = \Omega_0 \hat{\bm{e}}_\phi$} &
        \multicolumn{3}{c}{$\omega_\mathrm{A} \equiv \frac{B_0}{\sqrt{\bar{\rho}
        r}}$} &  \multicolumn{4}{c}{$N^2 \equiv -\frac{\partial\rho_0}{\partial
        z}\frac{g}{\bar{\rho}}$} \\ 
        \bottomrule[1pt]
        \end{tabular}
    \end{center}
    \caption{Parameters used in the simulations, with their corresponding name elements if applicable.}
    \label{tb:para}
\end{table*}

The non-dimensionalized parameters used in our simulations are summarized in
Tab. \ref{tb:para}. Since the simulations span a range of parameters (mainly the
rotation frequency $\Omega_0$, initial Alfv\'en frequency $\omega_\mathrm{A}$,
and Brunt--V\"{a}is\"{a}l\"{a} frequency $N$), a combination of these name
elements in Tab. \ref{tb:para} is used to refer to one simulation where a
certain combination of parameters are adopted, e.g., the notation ``{\tt
Om.5\_OmA.25\_N1}'' refers to the simulation with $\Omega_0 = 0.5$,
$\omega_\mathrm{A}=0.25$ and $N=1$. 

We note that like most other numerical work, our simulations depart from the
actual parameters due to limited computational power. The Reynolds number
$\mathrm{Re}\equiv 1/\nu$ and the magnetic Reynolds number $\mathrm{Re_m} \equiv
1/\eta$ used here are much smaller than those in real stars. In addition, real
RGB stars likely have $\omega_\mathrm{A} / \Omega \sim 10^{-1} \ll 1$, Froude
number $\Omega / N \sim 10^{-4} \ll 1$ and magnetic Prandtl number
$\mathrm{Pr_m} \equiv \nu/\eta \ll 1$; however, our simulations can only reach
much smaller scale separations with $\omega_\mathrm{A} / \Omega \sim \Omega / N
\sim 0.5$ and $\mathrm{Pr_m} \sim 0.5$.

The scale separation in our simulations is constrained due to the following
reasons: (1) since the vertical ($z$ direction in code setup) length scales of
the Tayler instability must satisfy $l_z \lesssim z (\Omega/N)$, $\Omega/N$ thus
cannot be too small otherwise $l_z$ cannot be resolved; (2) since the growth
rate of the Tayler instability roughly scales as $\omega_\mathrm{A}^2/\Omega$,
the initial $\omega_\mathrm{A}$ (or $B_0$) cannot be too small, otherwise the
growth of the Tayler instability will be too slow for the simulations to follow;
3) since the Tayler instability occurs when $\omega_\mathrm{A} / \Omega \gtrsim
(N/\Omega)^{1/2} (\eta/ r^2 \Omega)^{1/4}$ \citep{spruit:02,zahn:07}, given
$\omega_\mathrm{A} / \Omega < 1$ and $\Omega/N < 1$, the magnetic resistivity
$\eta$ needs to be small enough to allow the Taylor instability to develop, but
not too small to be unresolved on the grid scale. Similarly, although
$\mathrm{Pr_m} \equiv \nu/\eta \ll 1$ would be ideal, here we use $\mathrm{Pr_m}
= 0.5$ so that $\nu$ will not be too small to be resolved either.

We shall see that since the scale separation in our simulations
($\omega_\mathrm{A}\lesssim \Omega \lesssim N$ and $\mathrm{Pr_m} \lesssim 1$)
is much smaller than that in real stars ($\omega_\mathrm{A}\ll \Omega \ll N$ and
$\mathrm{Pr_m} \ll 1$), we do not expect the scaling relations measured from the
simulations to perfectly replicate theoretical predictions made under the limit
of large-scale separations (e.g., \citealt{fuller:19}). Nevertheless, our
simulations probe more realistic parameter spaces with the correct ordering of
scales $\omega_\mathrm{A} < \Omega < N$ that occur in real stars. This parameter
space has not been fully explored yet in previous studies, such as
\citet{braithwaite:06} who used $\Omega/N = 1$ and compressible fluid equations,
\citet{weber:15} and \citet{gellert:08}, who used $N=0$, and \citet{guerrero:19}
who used $\Omega=0$. As will be seen in the following sections, this setup
enables a clean and detailed numerical study of the Tayler instability and its
saturation while retaining much of the key physics for stellar interiors.

\section{Results}
\label{sec:results}

\subsection{Morphologies}

\begin{figure*}
  \begin{centering}
    \includegraphics[width=0.99\textwidth]{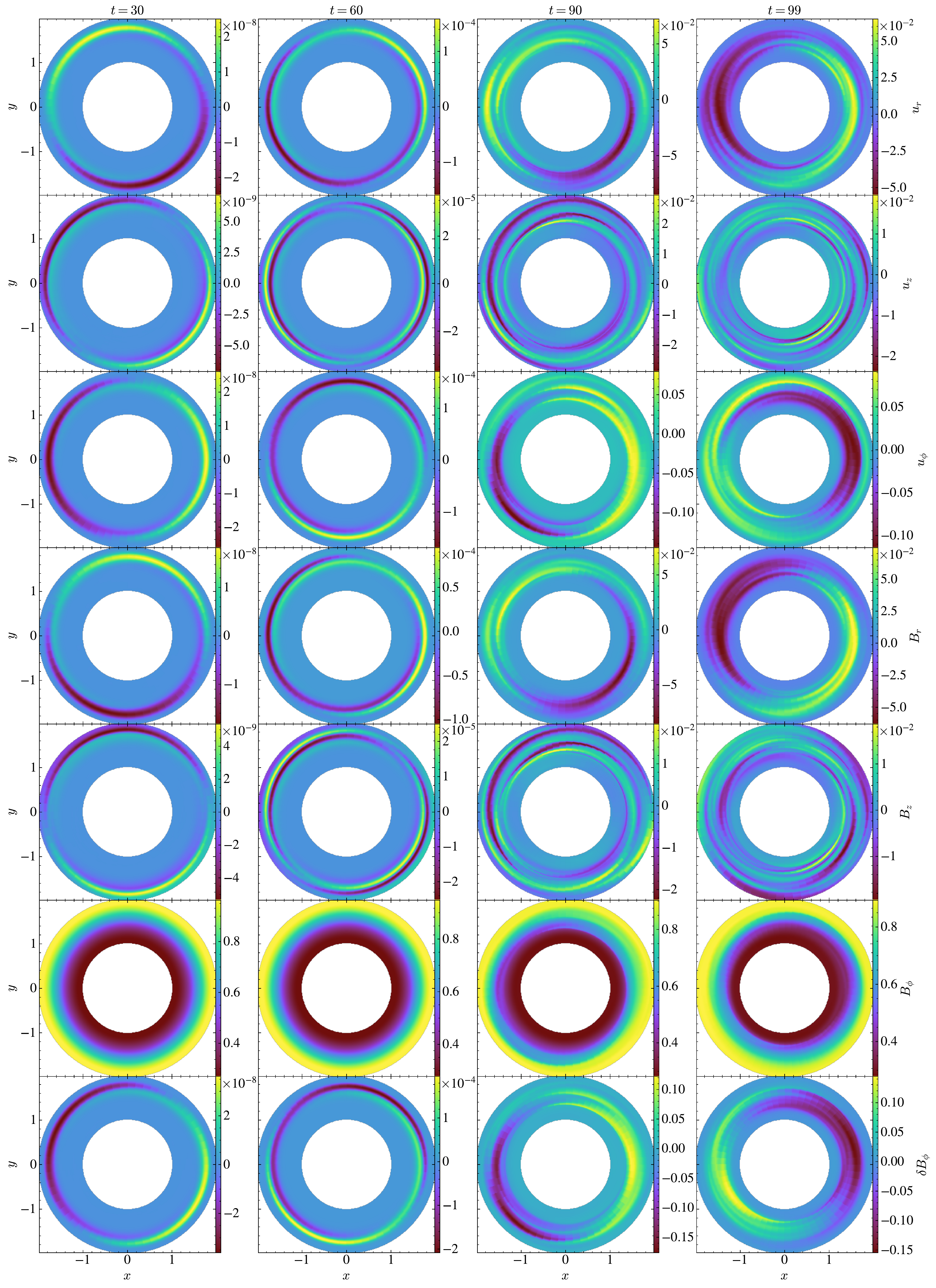}
  \end{centering}
  \vspace{-0.5cm}
  \caption{Slices at the mid-plane $z = 0$ of selected fluid quantities (from
  top to bottom): the velocity components $u_r$, $u_z$, $u_\phi$, the magnetic
  field components $B_r$, $B_z$, $B_\phi$ and the azimuthal magnetic field
  perturbations $\delta B_\phi$, at $t = 30, 60, 90$ and $99$ (from left to
  right) in the fiducial simulation {\tt Om.5\_OmA.25\_N1}. The early linear
  growth phase at $t\lesssim 60$ features an $m=1$ mode, with the amplitudes of
  fluid quantities growing exponentially. The $m=1$ mode is later mixed with
  higher modes after $t \! \gtrsim \!  80$, with the amplitudes reaching
  saturation. \label{fig:slice_z}\vspace{-0.2cm}}
\end{figure*}

\begin{figure*}
\begin{centering}
  \includegraphics[width=\textwidth]{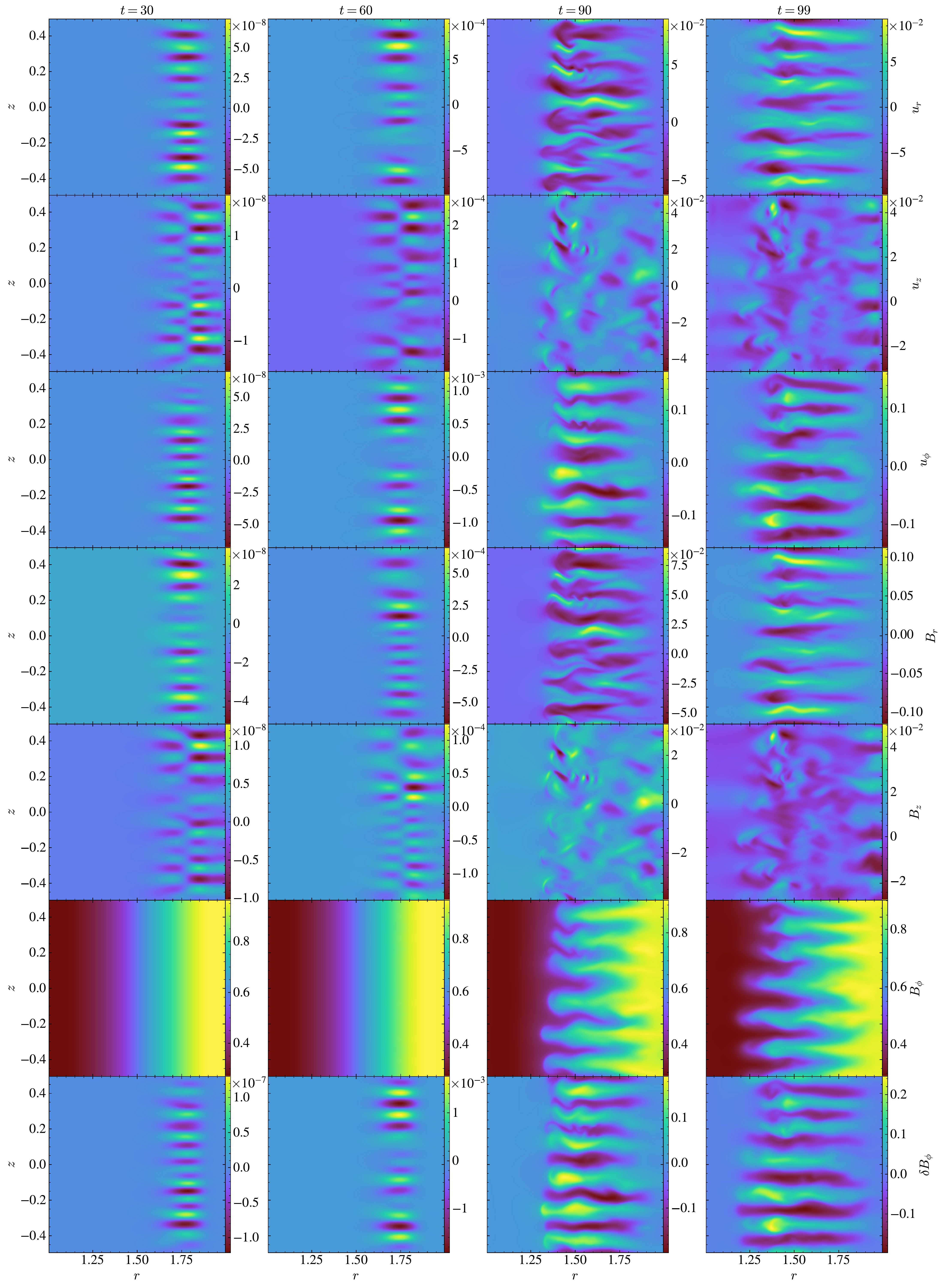}
\end{centering}
\vspace{-0.5cm}
\caption{Slices of the $r-z$ plane at $\phi = 0$, as Fig.~\ref{fig:slice_z}.
During the nonlinear growth and saturation phases at $t = 90$ and $99$,
horizontally aligned structures are strongly distorted and propagate toward
smaller radii. The $u_z$ and $B_z$ structures appear turbulent, indicative of a
secondary shear instability. \label{fig:slice_theta}\vspace{-0.2cm}}
\end{figure*}

We first study the run {\tt Om.5\_OmA.25\_N1} as a fiducial case.
Fig.~\ref{fig:slice_z} shows the time evolution at $t = 30, 60, 90$ and $99$
(from left to right columns) of each velocity component ($u_r$, $u_z$ and
$u_\phi$) and magnetic field component ($B_r$, $B_z$ and $B_\phi$), along with
the azimuthal magnetic field perturbations ($\delta B_\phi \equiv B_\phi -
\langle B_\phi \rangle_\phi$, where $\langle ...\rangle_\phi$ denotes azimuthal
averaging), in the $r-\phi$ plane. Note that the $m=1$ mode structures emerge by
$t = 30$ in Fig.~\ref{fig:slice_z}, as expected for Tayler instability. The
amplitudes of the $m = 1$ mode grow exponentially, up to $\sim 10^{-8}$ by
$t=30$ and $\sim 10^{-4}$ by $t=60$. By $t = 90$ and $99$ (right two columns of
Fig.~\ref{fig:slice_z}), the amplitudes have saturated, and the $m=1$ structure
has mixed together with higher $m$ modes.

Fig.~\ref{fig:slice_theta} shows the time evolution of the same set of
quantities as Fig.~\ref{fig:slice_z}, but viewed in the $r-z$ plane. The
structures on the $r$-$z$ plane are mostly horizontally aligned, with a short
wavelength in the $z-$direction. Because the dominant perturbations have $m=1$,
the values of $u_r$, $u_z$, etc., oscillate in time as the flow pattern
propagates. In the non-linear and saturation phases, the banded structure is
mostly maintained, but with a slightly longer wavelength in the $z$-direction and
distorted structure. However, the $z$-components of the flow, $u_z$ and $B_z$,
become highly turbulent during the non-linear saturation phase, losing the
banded structure. This appears to be the result of secondary shear instabilities
that develop as the instability saturates. The unstable eigenmodes are initially
confined to $r \! \gtrsim \! 1.5$. Simultaneously with saturation, the flow
pattern migrates inward, reaching nearly the inner boundary. We will discuss
the transition to the nonlinear stage and the shear instability in the following
sections.

\subsection{Linear growth rate}

\begin{figure}
  \begin{centering}
    \includegraphics[width=0.45\textwidth]{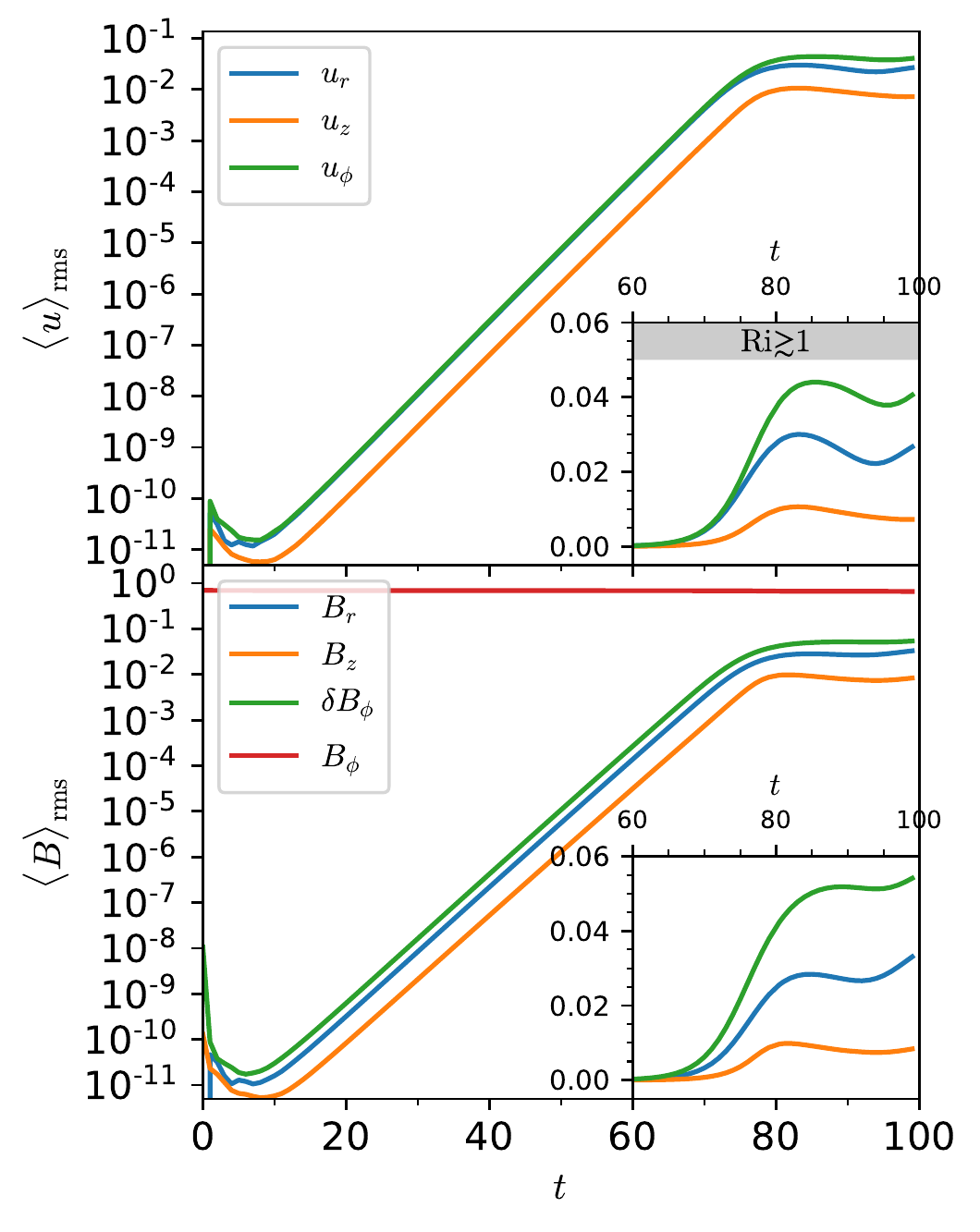}
  \end{centering}
  \vspace{-0.5cm}
  \caption{Time evolution of volume-averaged root-mean-squared velocities (top)
    and magnetic fields (bottom) in the fiducial simulation {\tt
    Om.5\_OmA.25\_N1}, with zoom-in on linear scales near the non-linear stage
    at $t > 60$. The regime where the Richardson number $\mathrm{Ri}\gtrsim1$
    (see \S\ref{sec:saturation}) is shadowed in the zoom-in plot. The magnitudes
    of both velocities and magnetic fields grow exponentially and finally
    saturate at $u_\perp = \sqrt{u_z^2 + u_\phi^2} \sim 0.05$ which corresponds
    to $\mathrm{Ri}\sim 1$. \label{fig:average}\vspace{-0.2cm}}
\end{figure}

\begin{figure}
  \begin{centering}
    \includegraphics[width=0.45\textwidth]{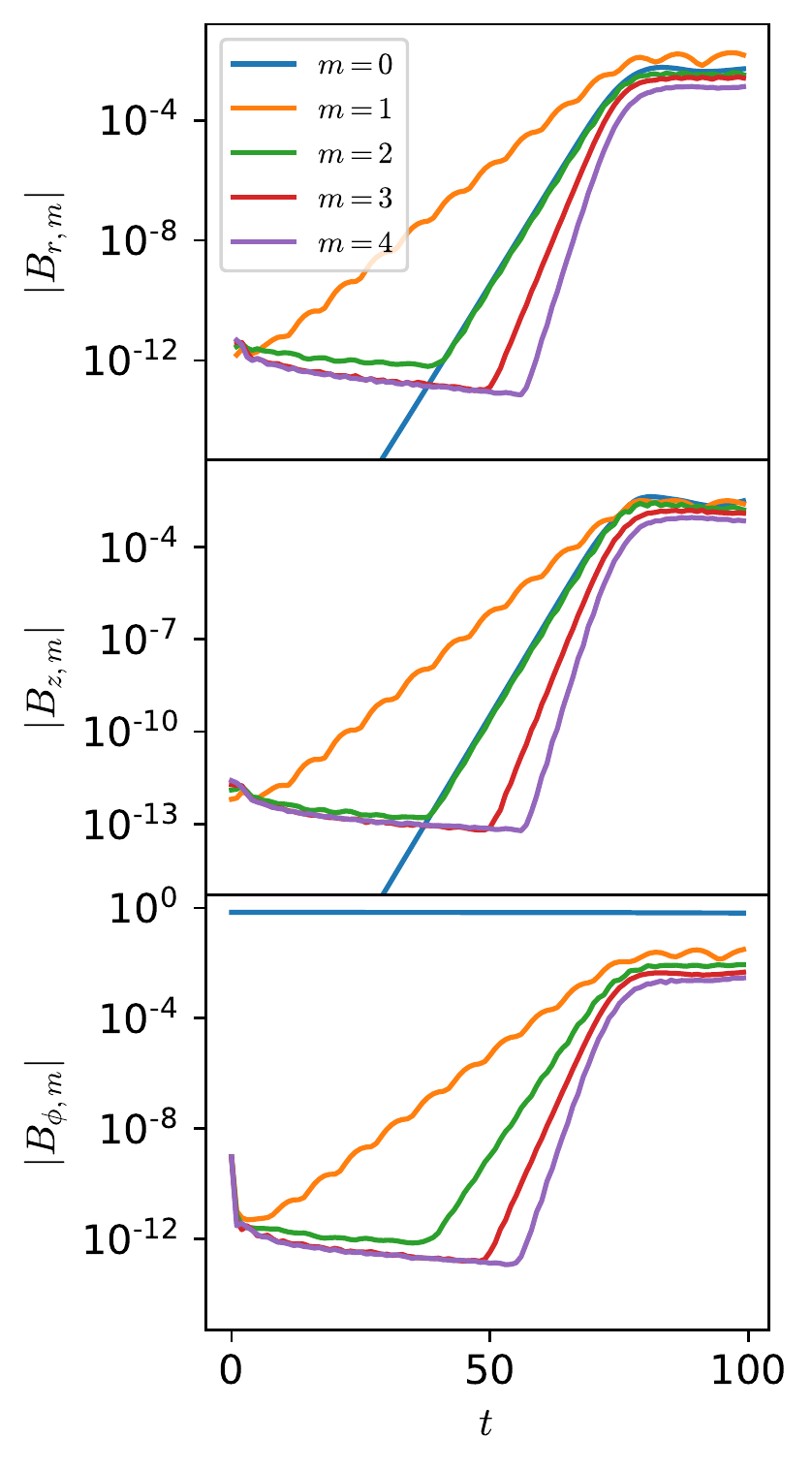}
  \end{centering}
  \vspace{-0.5cm}
  \caption{Time evolution of volume-averaged root-mean-squared azimuthal modes
    of magnetic fields $|B_m|$ in the fiducial simulation {\tt
    Om.5\_OmA.25\_N1}. The $m=1$ mode is dominant in the linear growth phase,
    and higher modes start to grow at later times, e.g., the $m = 0$ and $m = 2$
    mode grows from $t\sim 40$ at twice the rate of the $m = 1$ mode.
    \label{fig:B_mode_grow}\vspace{-0.2cm}}
\end{figure}

Fig.~\ref{fig:average} shows the time evolution of the root-mean-squared
velocities (top) and magnetic fields (bottom) in the fiducial simulation {\tt
Om.5\_OmA.25\_N1}. The evolution of both velocities and magnetic fields goes
through a linear growth stage for $t \sim 80$ before reaching saturation. As
expected, the magnitude of $u_z$ is smaller than $u_r$ and $u_\phi$ by a factor
of several, due to the buoyancy force that restricts motion in the
$z-$direction. Similarly, $B_z$ remains several times smaller than $B_r$ or
$\delta B_\phi$. At saturation, the perturbed field components remain more than
a factor of ten weaker than the background field $B_\phi$, which weakens only
slightly by the end of the simulation.

We further examine the growth of magnetic fields by decomposing them into
different azimuthal modes with the following equation
\begin{align}
  \left|B_m\right|^2
  \equiv \left\langle\left|\frac{1}{2\pi} \int  d\phi e^{i m \phi} B(r,z,\phi)\right|^2\right\rangle_{r,z}
\end{align}
with $m$ as azimuthal mode numbers, and $\langle...\rangle_{r,z}$ denoting
averaging over $r$ and $z$ under cylindrical coordinates. We plot the time
evolution of the amplitudes of different modes in Fig.~\ref{fig:B_mode_grow}.
The $m=1$ mode is dominant over higher $m$ modes in the linear growth phase,
consistent with the apparent $m = 1$ mode structure in Fig.~\ref{fig:slice_z}.
Modes of $m \geq 2$ start to grow at $t \sim 40$, and all modes ultimately reach
saturation at $t \sim 80$. We will discuss the nonlinear coupling and saturation
in \S\ref{sec:nonlinear} and \S\ref{sec:saturation}.

\begin{figure*}
  \begin{centering}
    \includegraphics[width=\textwidth]{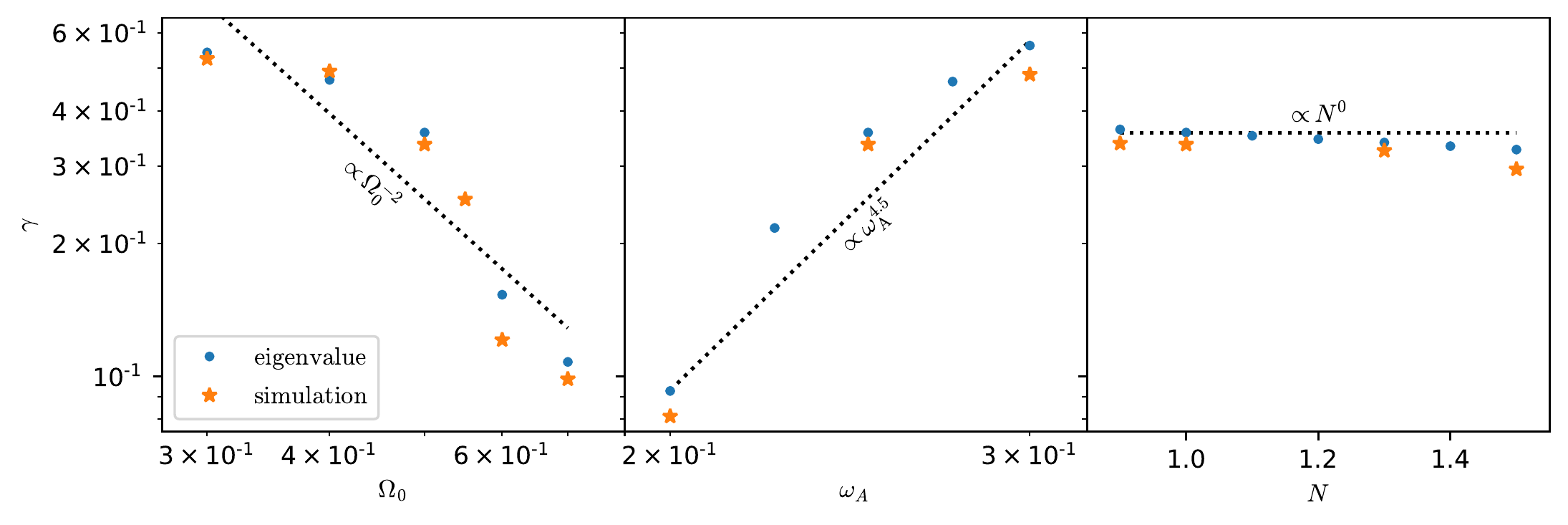}
  \end{centering}
  \vspace{-0.5cm}
  \caption{Linear growth rates $\gamma$ of the Tayler instability vs $\Omega_0$,
    $\omega_\mathrm{A}$ and $N$, given by linear eigenvalue calculations (blue
    dots) and measured from simulations (orange stars) with varying parameters.
    The linear growth rates in simulations are well-predicted by the linear
    eigenvalue calculations. \label{fig:lin_growth}\vspace{-0.2cm}}
\end{figure*}

\begin{figure}
  \begin{centering}
    \includegraphics[width=0.4\textwidth]{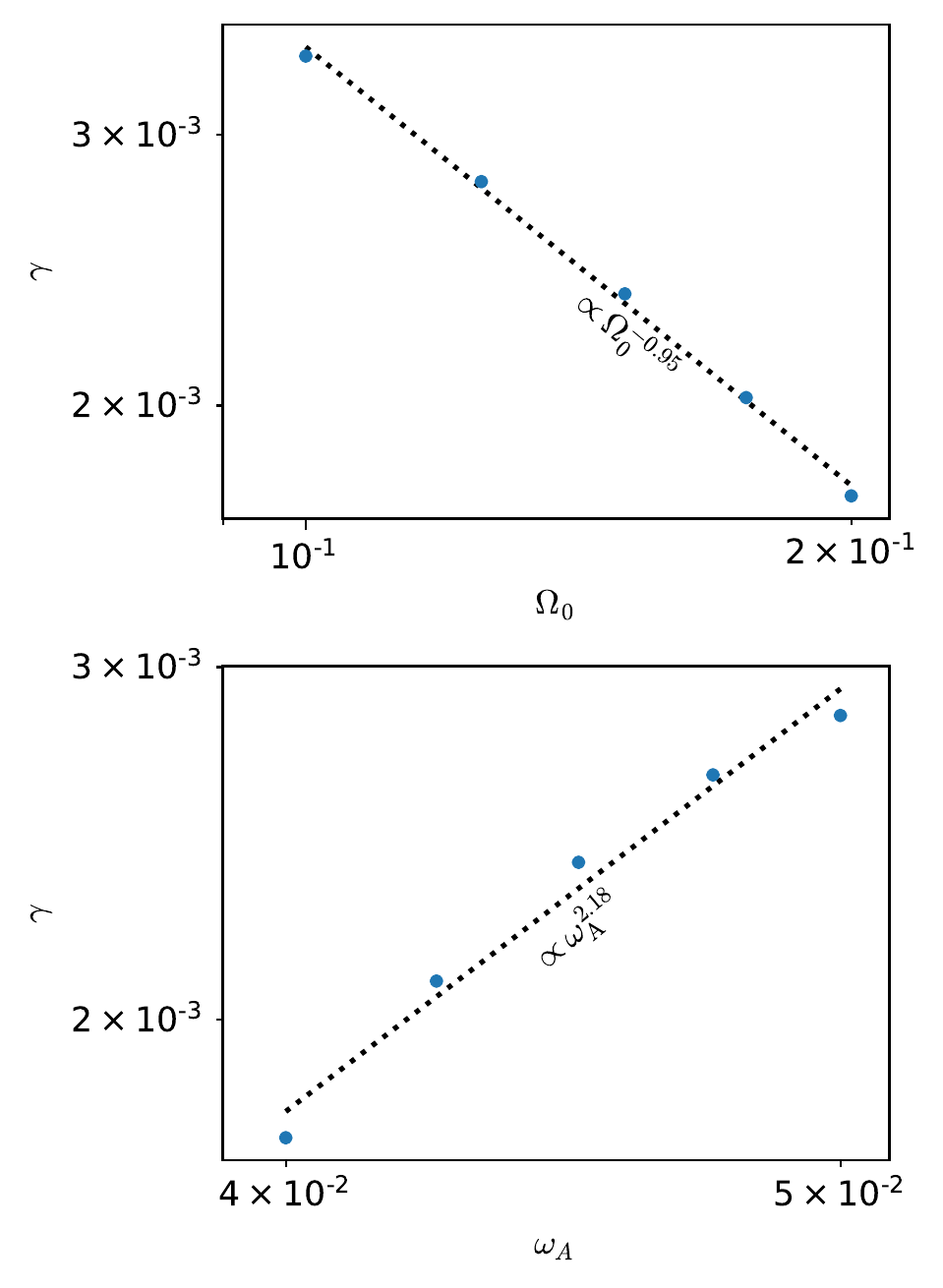}
  \end{centering}
  \vspace{-0.5cm}
  \caption{Linear growth rates $\gamma$ of the Tayler instability from linear
  eigenvalue calculations (blue dots), with much greater scale separations (
  $\omega_A \sim 0.04$ -- $0.05$, $\Omega_0 \sim 0.1$ -- $0.2$, $N = 1$ and
  $\eta = 5 \times 10^{-6}$) and consequently much smaller growth rates
  ($\gamma\sim$ a few $\times 10^{-3}$) than the fiducial simulations. With
  larger scale separations between $\omega_\mathrm{A}$, $\Omega_0$ and $N$, the
  best power-law fits (dashed line) is roughly consistent with the theoretical
  prediction of $\gamma \propto \omega_\mathrm{A}^2 \Omega_0^{-1}$
  \citep{spruit:99}, and the growth rates are much smaller than those in
  Fig.~\ref{fig:lin_growth}. \label{fig:lin_growth_large}\vspace{-0.2cm}}
\end{figure}

We measure the linear growth rates of the Tayler instability from simulations
with varying initial-conditions parameters, including the angular frequency
$\Omega_0$, Aflv\'{e}n frequency $\omega_\mathrm{A}$ and
Brunt-V\"{a}is\"{a}l\"{a} frequency $N$. The growth rates measured from
simulations (orange stars) are plotted against the eigenmodes (blue dots) of the
Tayler instability setup calculated with the eigenvalue problem solver in
{\small Dedalus}. The simulated linear growth rates $\gamma$ are well-predicted
by the linear eigenvalue calculations, following a scaling relation of $\gamma
\propto \omega_\mathrm{A}^{4.5} \Omega_0^{-2} N^0$ (Fig. \ref{fig:lin_growth}).
This scaling is different than expected from \citet{spruit:99} (see also
\citealt{zahn:07} and \citealt{ma:19}), who predicts the fastest growing modes
scale as $\gamma \sim \omega_\mathrm{A}^2 \Omega_0^{-1}$ in the limit that
$\omega_\mathrm{A} \ll \Omega_0$. This occurs because our simulations are not
actually in the asymptotic limit of $\omega_{\rm A} \ll \Omega \ll N$ used for
the analytic estimates in those works: in Fig.~\ref{fig:lin_growth_large}, we
further carry out linear eigenvalue calculations with larger scale separations
with $\omega_A \sim 0.04$ -- $0.05$, $\Omega_0 \sim 0.1$ -- $0.2$, $N = 1$ and
$\eta = 5 \times 10^{-6}$, and find that the obtained scaling relations of the
growth rates are quite consistent with the predictions by \citet{spruit:99}.
However, the resulting growth rates are as low as a few $10^{-3}$, which are
much smaller than those with fiducial parameters ($\gamma \sim$ a few $\times
10^{-1}$ in Fig.~\ref{fig:lin_growth}) and are prohibitively small for numerical
simulations to follow the growth of the Tayler instability. Therefore, we stick
to the fiducial parameters for the simulations even though they have limited
scale separations, and bear it in mind when comparing our results with analytic
estimates that the simulations are not fully in the limit of $\omega_{\rm A} \ll
\Omega_0$ as used in many analytic works.

\subsection{Nonlinear coupling}
\label{sec:nonlinear}

From Fig.~\ref{fig:B_mode_grow}, we can see that during the linear growth phase,
the amplitude of the $m=1$ mode grows exponentially as expected. The $m=0$ and
$m\geq2$ modes initially decay because they are stable, but eventually they also
start growing exponentially. This is a consequence of non-linear power transfer
from the large amplitude $m=1$ mode to other values of $m$. From the induction
equation \eqref{eq:induction}, we can see that the magnetic field grows as
(neglecting diffusive effects which are small in this analysis)
\begin{equation}
    \frac{\partial \bm{B}}{\partial t} 
        \simeq \nabla\times\left(\bm{u}\times\bm{B}\right) \, .
\end{equation}
During the linear growth stage, fluctuations in $\bm{u}$ and ${\bm B}$ are
dominated by the $m=1$ component, which has time and spatial dependence $B
\propto \sin (\phi - \omega t) e^{\gamma t}$, and similar for $u$. Here $\omega$
is the real part of the frequency of the fastest growing mode, and $\gamma$ is
its growth rate. Hence, to lowest order, the $m \neq 1$ components grow as
\begin{align}
\label{eq:Bnonlin}
    \frac{\partial B}{\partial t} 
        &\propto \sin^2(\phi - \omega t) e^{2 \gamma t} \nonumber \\
        &\propto \frac{1}{2} \big[ 1 - \cos(2\phi - 2\omega t)\big] e^{2 \gamma t} \, .
\end{align}
for $m=0$ and $m \geq 2$ modes.

Multiplying each side of Eq.~\eqref{eq:Bnonlin} by $\sin(m \phi)$ and
integrating over volume gives the contribution to the non-linear growth of
$\vec{B}$ at a desired wavenumber $m$. We see that to the lowest order, only the
$m=0$ and $m=2$ modes have a non-vanishing integral, arising from the first and
second terms inside the parentheses in Eq.~\eqref{eq:Bnonlin}, respectively.
Hence, we see that the $m=0$ and $m=2$ modes grow at exactly twice the rate as
the $m=1$ mode, as long as the $m=1$ mode has much larger amplitude. At a given
time $t$, the amplitude of the non-linearly excited modes scales as $\vec{B}
\propto \int \dot{\vec{B}} dt \propto B(r,z,m=1,t=0)^2 e^{2 \gamma t}$. Hence at
a given time $t$, the ratio of the non-linearly excited mode to the linearly
excited mode is 
\begin{equation}
 \frac{|B(m=0,2)|}{|B(m=1)|} = \kappa B(m=1) \, ,
\end{equation}
where $\kappa$ is a constant of proportionality that captures the non-linear
coupling between modes. From Fig.~\ref{fig:B_mode_grow}, we estimate $\kappa
\sim 0.1/B_\phi$, where $B_\phi$ is the background magnetic field strength.
Hence,  the non-linearly excited modes have much smaller amplitudes in the
linear regime where $B(m=1) \ll B_\phi$.

Extending this calculation to $m>2$ requires higher chains of non-linear
interaction that results in a non-linear growth rate $m \gamma$ for modes with
$m \geq 2$. This scaling is verified by the growth rates and amplitudes shown in
Fig.~\ref{fig:B_mode_grow} in the linear regime, with $t \lesssim 70$. Hence, in
the linear growth phase, we clearly see a non-linear transfer of power to
smaller scales. By the time the instability saturates, the high-$m$ modes have
reached amplitudes comparable to (but smaller than) the $m=1$ mode, and the
weakly non-linear analysis presented above begins to break down.

\subsection{Non-linear saturation}
\label{sec:saturation}

The end of linear growth and apparent saturation of the instability at around
$t=80$ in our simulations is accompanied by a remarkable change in the motions
and morphology of the simulation domain (see Fig.~\ref{fig:slice_theta}). This
includes the appearance of turbulent non-layered structure in $u_z$ and $B_z$
(Fig.~\ref{fig:slice_theta}), as well as the inward motion of the perturbed flow.

The turbulent structure appears to stem from secondary shear instabilities. We
believe the evolution is similar to the magnetized Rayleigh-Taylor instability,
where the primary instability also drives opposing flows which then exhibit
shear instabilities \citep[e.g.,][]{stone:07}. For the Tayler instability as we
consider here, buoyancy provides a restoring force in the $z$ direction. The
horizontal flows driven by the instability will become unstable to secondary
Kelvin-Helmholtz instabilities when
\begin{equation}
\label{kh}
k_z u_\perp \gtrsim 2 N \, ,
\end{equation}
where $u_\perp = \sqrt{ u_r^2 + u_\phi^2}$ is the fluid velocity perpendicular
to $\hat{z}$. Near the end of the simulation, the dominant modes have roughly
seven wavelengths in the $z-$direction and hence $k_z \sim 40$. Therefore in our
simulations with $N=1$, we expect shear instabilities to occur when $u_\perp
\sim u_\phi \gtrsim 0.05$. Indeed, Fig.~\ref{fig:average} shows that the system
reaches its saturated state very near this scale, and that this saturation is
accompanied by turbulence shown in Fig.~\ref{fig:slice_theta}. Despite the shear
instabilities, $u_r$, $u_\phi$, $B_r$, and $\delta B_\phi$ all maintain a banded
structure in the $z$-direction, and they also maintain a predominantly $m=1$
azimuthal structure.

At the same time that secondary instabilities develop, the unstable region also
``migrates inward". During linear growth, the instability is restricted to large
radii with $r \gtrsim 1.5$, because the Tayler instability only occurs in the
outer part of the domain where the background magnetic field is stronger. As the
instability saturates around $t=80$, however, the unstable motions and magnetic
field perturbations move inwards to the inner boundary at $r=1$. The inward-moving fingers have a predominantly $m=1$ azimuthal structure and feature inward
radial motion at alternating heights, accompanied by an outward return flow in
between. Their structures resemble convective plumes/fingers seen in the early
phases of convective/thermohaline instability.

We posit that the secondary instabilities cause saturation both by dissipating
energy from the growing modes, and by allowing for the inward motion of the
instability. Once the fluid elements obtain velocities large enough to overcome
the Richardson criterion of Eq.~\eqref{kh}, inertial forces become comparable to
buoyancy forces, and fluid elements can flow with fewer constrictions. This may
allow for circulation to smaller radii that was previously prevented by
buoyancy/Coriolis forces. It is also possible that non-linear inertial terms
effectively change the linear dispersion relation, allowing unstable motions to
grow at smaller radii. Future work should investigate this effect in detail, and
determine whether latitudinal migration of the Tayler instability could occur in
real stars.

\subsection{Dynamo action}
\label{sec:dynamo}

A major question regarding the Tayler-Spruit dynamo is whether a dynamo actually
occurs. In the picture advanced by \citet{spruit:02}, differential rotation
amplifies $B_\phi$ by winding up a weak poloidal field. The dynamo loop is
supposed to be closed by the amplification of the poloidal field via the
unstable motions associated with Tayler instability. However, as pointed out by
\citet{zahn:07} (see also \citealt{fuller:19}), the unstable motions are $m=1$
and do not directly produce an axisymmetric component of the poloidal field that
is needed for amplification of $B_\phi$ via differential rotation. Hence, a
non-linear coupling process is needed to amplify the axisymmetric poloidal
field, and this step of the dynamo is poorly understood.

\begin{figure}
  \begin{centering}
    \includegraphics[width=0.45\textwidth]{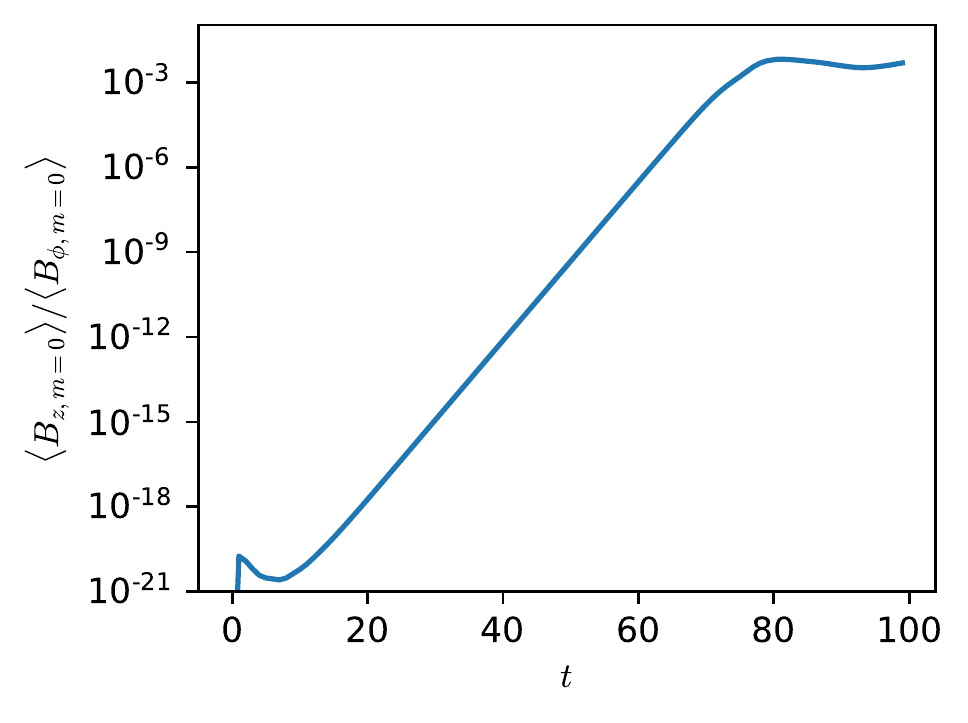}
  \end{centering}
  \vspace{-0.5cm}
  \caption{Time evolution of the ratio of axisymmetric ($m=0$ mode) $B_z$ to
  $B_\phi$ components from the simulation {\tt Om.5\_OmA.25\_N1}, demonstrating
  the non-linear induction process needed to amplify the axisymmetric poloidal
  field to close a dynamo loop. \label{fig:B_m0_vs_time}\vspace{-0.2cm}}
\end{figure}

Fig.~\ref{fig:B_m0_vs_time} shows the growth of the axisymmetric component of
$B_z$, $B_{z,m=0}$, which is the poloidal component of the field needed to be
generated by the Tayler instability in order for a dynamo to occur. Our
simulations do not include differential rotation and thus cannot capture the
winding of $B_{z,m=0}$ needed to close the dynamo loop. However, our simulations
clearly demonstrate non-linear amplification of $B_{z,m=0}$, as also discussed
in Section \ref{sec:nonlinear}. We have verified that the proportionality of the
scaling $B_z/B_\phi \sim \omega_{\rm A}/N$ predicted by \citet{spruit:02} and
\citet{fuller:19} roughly holds, but with a much smaller normalization factor of
$B_z \sim 10^{-2} B_\phi$ for the saturated state of our simulations, rather
than the predicted $B_z \sim (\omega_{\rm A}/N) B_\phi \sim 1/4 B_\phi$, i.e.,
the axisymmetric component of $B_z$ in our simulations saturates at much lower
values than predicted by the scaling relation. However, both of those models are
based on dynamos sustained by energy input by shear, which cannot occur in our
simulations, so the disagreement is not surprising.

Future work will be necessary to understand the saturated state of the
Tayler-Spruit dynamo, but our simulations do indicate that a dynamo based on
non-linear induction can greatly amplify the axisymmetric component of the
poloidal field.

\subsection{Scaling Relations}

\begin{figure*}
  \begin{centering}
    \includegraphics[width=\textwidth]{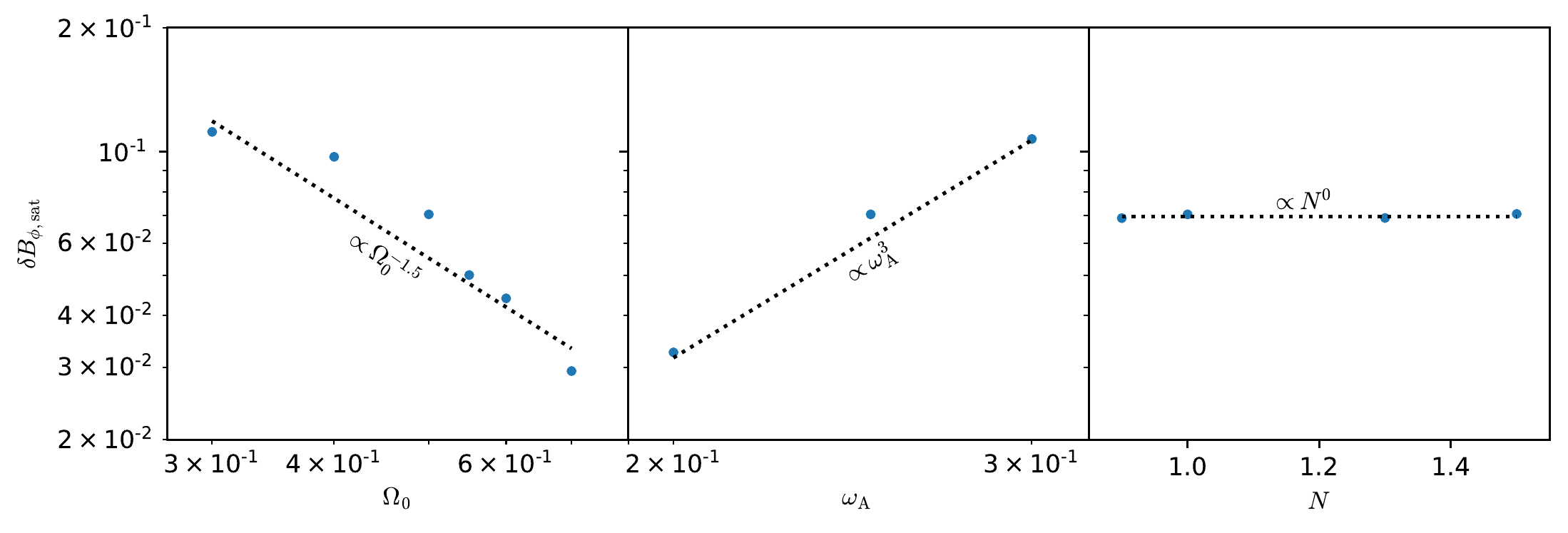}
  \end{centering}
  \vspace{-0.5cm}
  \caption{Saturated RMS values of $\delta B_\phi$ as a function of $\Omega_0$,
    $\omega_\mathrm{A}$ and $N$, measured from simulations with varying
    parameters at the saturation stage. The saturated magnitudes of $\delta
    B_\phi$ strongly correlate with the $\Omega_0$ and $\omega_\mathrm{A}$, and
    they are approximately independent of $N$ as expected.
    \label{fig:saturation}\vspace{-0.2cm}}
\end{figure*}

A key result of our simulations is how the properties of the saturated state
depend on input parameters ($\omega_{\rm A}$, $\Omega_0$, etc.). Fig.
\ref{fig:saturation} shows how the mean amplitude of the magnetic field
perturbations in the saturated state, $\delta B_{\phi, {\rm sat}}$, scale with
$\Omega_0$, $\omega_{\rm A}$, and $N$. The saturated values of $\delta B_{\phi,
{\rm sat}}$ scale strongly with rotation rate and initial magnetic field, with
$\delta B_{\phi, {\rm sat}} \propto \Omega_0^{-1.5}$ and $\delta B_{\phi, {\rm
sat}} \propto \omega_{\rm A}^3$, and almost no dependence on $N$. These scalings
are similar to (but slightly weaker than) the linear growth rate scaling
presented in Fig. \ref{fig:lin_growth}.

Based on the discussion above, we believe the instability saturates in our
simulations due to secondary shear instabilities. For the large-scale growing
modes with $m\sim 1$, the non-linear inertial term $(\bm{u} \cdot \nabla) {\bm
u}$ produces an effective damping rate of order $\gamma_{\rm NL} \sim 2 \pi
u_\phi/r$. In the rapidly rotating limit, we expect $ u_\phi  \propto
(\omega_{\rm A}/\Omega_0) \delta B_{\phi, {\rm sat}}$ \citep{spruit:02,
fuller:19}, so the non-linear damping rate will scale as $\gamma_{\rm NL}
\propto \delta B_{\phi, {\rm sat}} (\omega_{\rm A}/\Omega)$. Setting this equal
to the linear growth rate, which scales as $\gamma \propto \omega_{\rm
A}^{4.5}/\Omega_0^2$ (Fig. \ref{fig:lin_growth}) yields an expected scaling of
the non-linear saturation amplitude of $\delta B_{\phi, {\rm sat}} \propto
\omega_{\rm A}^{3.5}/\Omega_0$, similar to the scaling Fig.
\ref{fig:saturation}. The slight mismatch between these scaling laws could stem
from the small dynamic range covered by our simulations, or the fact that they
are not quite in the rapidly rotating limit. 

\citet{fuller:19} argued that Tayler instability saturates via weak magnetic
turbulence with a non-linear damping rate of $\gamma_{\rm NL} \sim \delta
\omega_{\rm A} \propto \delta B_{\phi, {\rm sat}}$. Setting this expectation
equal to the linear growth rates would entail $\delta B_{\phi, {\rm sat}}
\propto \omega_{\rm A}^{4.5}/\Omega_0^2$, steeper than the scalings shown in
Fig. \ref{fig:saturation}. Hence, it appears that our simulations do not
saturate via weak magnetic turbulence as predicted by \citet{fuller:19}.
However, in the next section, we discuss why the saturation mechanism is likely
to be different when the Tayler instability operates in stellar interiors.

\section{Discussion and Conclusion}
\label{sec:diss}

Although secondary shear instabilities appear to saturate the Tayler instability
in our simulations, it is likely that a different mechanism will operate in real
stars. \citet{fuller:19} predict that non-linear Alfv\'en wave dissipation
produces damping at the rate $\gamma_{\rm NL} \sim \delta v_{\rm A}/r$, which in
our units equates to $\gamma_{\rm NL} \sim \delta B_\phi/r \sim 0.04$ in the
saturated state. The dimensionless growth rate (Fig.~\ref{fig:lin_growth}) is
larger than this, roughly 0.4 for the fiducial simulation {\tt
Om.5\_OmA.25\_N1}. This indicates that the instability saturates via shear
instabilities before it reaches an amplitude high enough for Alfv\'en wave
dissipation to dominate. In a real star, however, Alfv\'en wave dissipation
would likely occur before shear instability. As described above, the dissipation
rate from shear instabilities is $\gamma_{\rm NL} \sim 2 \pi u_\phi/r$, and we
expect $u_\phi \sim (\omega_{\rm A}/\Omega_0) \delta v_{\rm A}$ in the limit
$\omega_{\rm A}/\Omega_0 \ll 1$ applicable to real stars. Because our
simulations only reach $\omega_{\rm A}/\Omega_0 \sim 1/2$, dissipation via shear
instability and Alfv\'enic dissipation are comparable, and evidently, shear
instability is more important by a factor of a few. But in a real star with
$\omega_{\rm A}/\Omega_0 \ll 1$, shear instabilities become less important, and
Alfv\'enic dissipation is expected to dominate. Hence, our simulations cannot
grow to an amplitude high enough to test the prediction of \citet{fuller:19},
because of the small ratio of $\Omega$ to $\omega_{\rm A}$. Future work should
push towards larger scale separations to better test these predictions.

\begin{figure}
  \begin{centering}
    \includegraphics[width=0.45\textwidth]{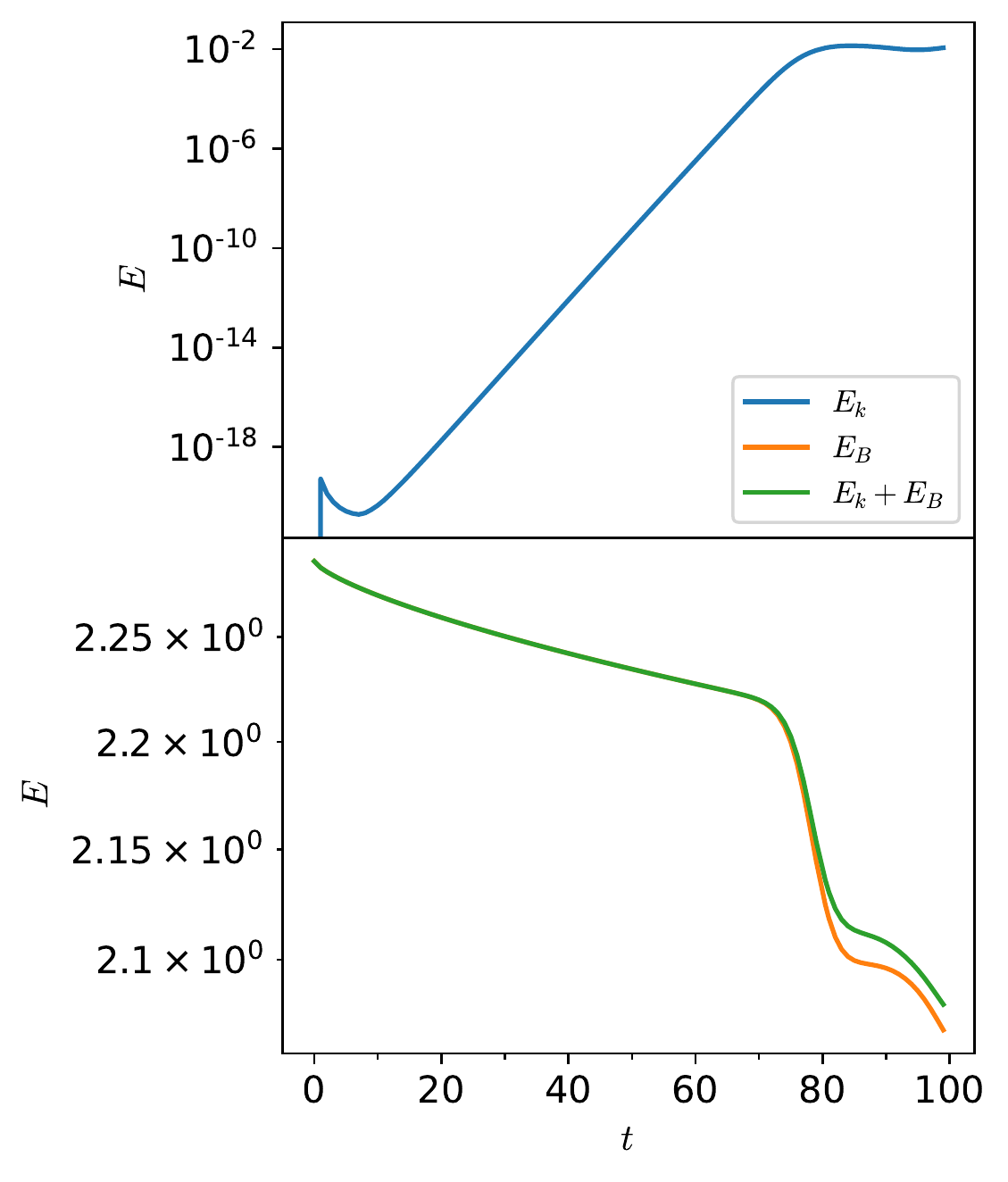}
  \end{centering}
  \vspace{-0.5cm}
  \caption{Time evolution of the volume-integrated kinetic $E_k$ (blue),
  magnetic $E_B$ (orange) and total $E_k + E_B$ (green) energies from the
  fiducial simulation {\tt Om.5\_OmA.25\_N1}. Together with the
  order-of-magnitude growth of the kinetic energy driven by the Tayler
  instability, the magnetic energy slightly dissipates at a rate roughly
  consistent with the prediction of $\sim \gamma |\delta B_\phi|$, where
  $\gamma$ is the growth rate of the Tayler instability.
  \label{fig:E_vs_time}\vspace{-0.2cm}}
\end{figure}

To determine whether Alfv\'en wave damping or shear instability will saturate
Tayler instability, we can determine which mechanism requires a lower threshold
to operate. As described above, shear instability requires $u_\perp \! \sim \! 2
N/k_z$. The fastest growing modes have $k_z^2 \eta \! \sim \! \omega_{\rm
A}^2/\Omega$, so shear instability operates when $u_\perp \! \sim 2 (\Omega
\eta)^{1/2} N/\omega_{\rm A}$. In contrast, Alfv\'en wave damping saturates the
instability when $\delta v_{\rm A} \! \sim \! r \omega_{\rm A}^2/\Omega$. Using
$\delta v_{\rm A} \! \sim \! (\Omega/\omega_{\rm A}) u_\perp$, the corresponding
velocity amplitude for Alfv\'en wave damping is $u_\perp \! \sim \! r
\omega_{\rm A}^3/\Omega^2$. Hence, Alfv\'en wave damping will dominate when $r
\omega_{\rm A}^3/\Omega^2 \! \lesssim \! 2 (\Omega \eta)^{1/2} N/\omega_{\rm
A}$, which translates to $\omega_{\rm A}/\Omega \! \lesssim \! (N/\Omega)^{1/4}
(\eta/r^2 \Omega)^{1/8} \equiv x$. We note that the instability growth criterion
is $\omega_{\rm A}/\Omega \! \gtrsim \! (N/\Omega)^{1/2} (\eta/r^2 \Omega)^{1/4}
= x^2$. Hence, Alfv\'en wave damping is expected to cause saturation when $x^2
\! \lesssim \! \omega_{\rm A}/\Omega \! \lesssim \! x$, while shear instability
is expected to cause saturation when $\omega_{\rm A}/\Omega \! \gtrsim \! x$.
Our simulations have $\omega_{\rm A}/\Omega \! \approx \! 0.5$, and $x \!
\approx \! 0.3$, hence we expect them to saturate via shear instability.
However, simulations that reach lower values of $\omega_{\rm A}/\Omega$ are
expected to exhibit saturation of the instability via Alfv\'en wave damping. We
hope to explore this possibility in future work.

Another argument against shear instability being important in real stars is as
follows. According to the saturated state of \citet{fuller:19}, the value of
$k_z u_\perp$ reaches $k_z u_\perp \sim q \Omega^2/N$, where $q = d \ln \Omega/d
\ln z$ is the dimensionless shear. This is much smaller than $N$ unless $q
\Omega^2 \gtrsim N^2$, which cannot occur in real stars, since a background
state with $q \Omega^2 \gtrsim N^2$ would already satisfy the Richardson
criterion for overturning the stratification. Therefore, at the saturated state
due to Alfv\'en wave dissipation, the value of $k_z u_\perp$ would be much less
than required to overturn the stratification and cause shear instability.
Simulations including shear and pushing to smaller ratios of $\Omega_0/N$ will
be needed to test this prediction.

\begin{figure}
  \begin{centering}
    \includegraphics[width=0.45\textwidth]{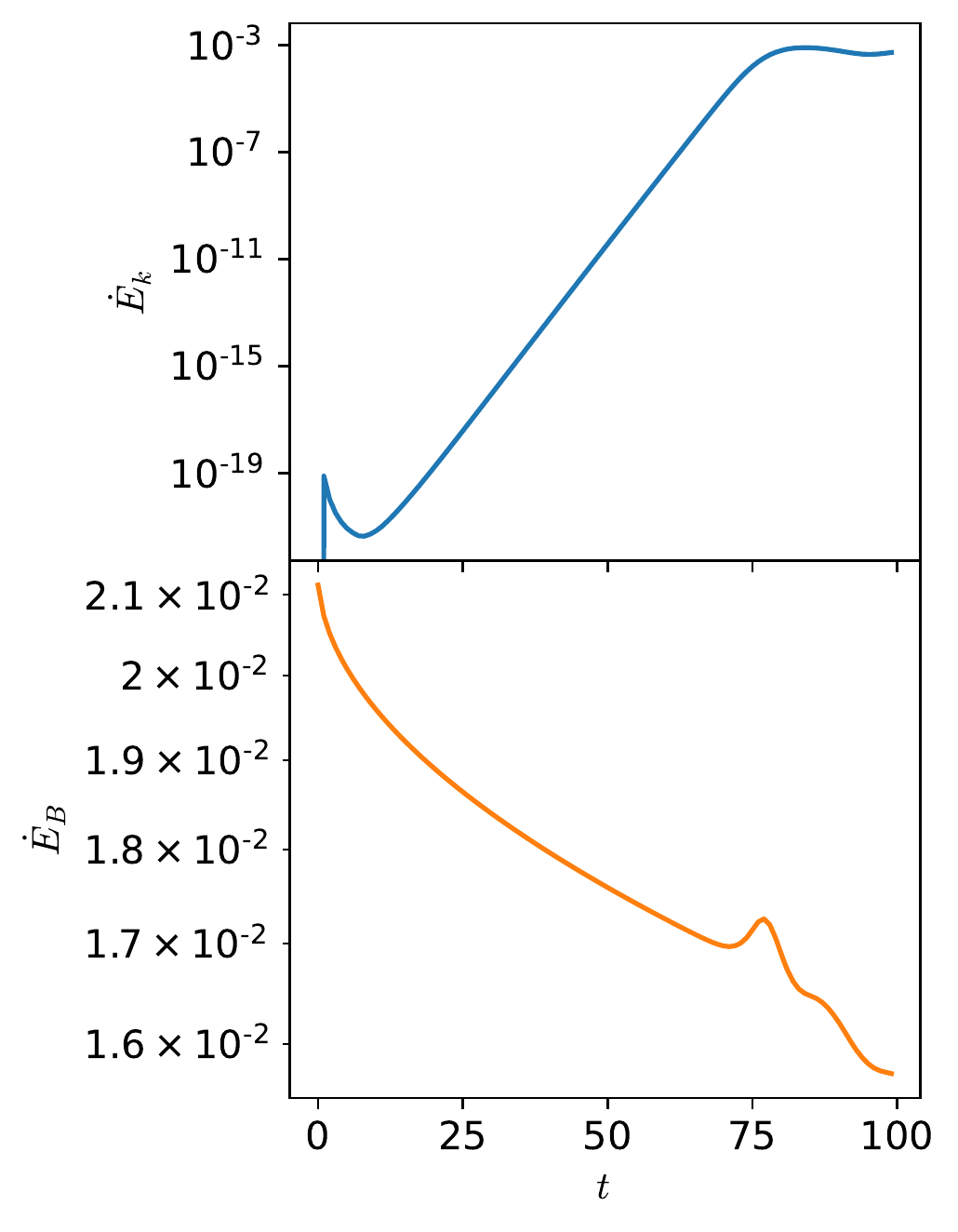}
  \end{centering}
  \vspace{-0.5cm}
  \caption{Time evolution of the volume-integrated kinetic $\dot{E}_k$ (top) and
  magnetic $\dot{E}_B$ (bottom) energy dissipation rates from the fiducial
  simulation {\tt Om.5\_OmA.25\_N1}.
  \label{fig:dissipation_vs_time}\vspace{-0.2cm}}
\end{figure}

Our simulations appear to disagree with the model of \citet{spruit:02}, in which turbulent dissipation in the saturated state dissipates magnetic energy at the rate $\dot{E} \sim \gamma B_\phi^2$, where $\gamma$ is the linear growth rate (equal to the non-linear damping rate) of the Tayler instability. Even though our saturated state is somewhat turbulent, it does not dissipate magnetic energy at this rate, as can be seen from the bottom panel of Fig.~\ref{fig:E_vs_time}. Since the growth rate of the instability is $\gamma \sim 0.4$, this would imply that all of the magnetic energy would dissipate within a time span of $\Delta t \sim 2.5$. In contrast, Fig.~\ref{fig:E_vs_time} shows that the magnetic energy only decreases by $\sim$10\% in the final $\Delta t =20$ of our simulation, corresponding to a magnetic energy dissipation rate of $\dot{E}_B \sim 2 \times 10^{-2}$ as shown in Fig.~\ref{fig:dissipation_vs_time}. This corresponds to magnetic energy dissipation $\sim$2 orders of magnitude slower than $\dot{E} \sim \gamma B_\phi^2$. Fig.~\ref{fig:dissipation_vs_time} also shows that kinetic energy dissipation is negligible relative to magnetic energy dissipation. However, we note that our imposed background field $B_\phi$ is ordered (uniform in the $z-$direction), and a field built up by a dynamo could be disordered, allowing for more magnetic energy dissipation. Also, our boundary conditions prevent $B_\phi$ from being totally erased, which may limit the turbulent dissipation of $B_\phi$. Future work with different initial conditions and boundary conditions will shed light on this issue.

Shortly before this paper was submitted, \citet{petitdemange:23} presented a
suite of dynamo simulations driven by differential rotation and exhibiting the
Tayler instability in a spherical shell. Since
their methods were very different from ours, we do not attempt a direct
comparison with our results. In their setup, shear was created by enforcing the
outer boundaries to rotate at different rates, though this also created Ekman
boundary layer effects and a related ``weak dynamo" that complicated the
analysis. The Tayler instability was triggered once the magnetic field generated
by the weak dynamo exceeded the threshold derived from linear theory. The
saturation mechanism of the Tayler instability was not clear from that work, but
the magnetic torques in the saturated state appeared to scale according to the
predictions of \citep{spruit:02}. However, most of those simulations appeared to
have $\Omega/N \sim 1$\footnote{For the simulation that did have $\Omega/N \ll
1$, this was true at the outer boundary but not necessarily in the bulk of the
simulation where the Tayler instability occurred. Additionally, the resulting
magnetic torque deviated from the trend exhibited by the rest of the models.},
where the predictions of \citep{spruit:02} and \citet{fuller:19} are similar.
While the work of \citet{petitdemange:23} is a great leap forward in modeling
the Tayler instability, our understanding of the saturation mechanism and
magnetic torques in real stellar interiors remains incomplete.

\subsection{Conclusion}
\label{sec:conclusion}

In this study, we perform three-dimensional MHD simulations of the Tayler
instability in a cylindrical annulus, with strong buoyancy and Coriolis forces
incorporated to mimic realistic stellar environments. The simulations are
initialized with a strong toroidal field which is in magnetostatic equilibrium.
We explore a range of parameter space by varying the angular velocities
$\Omega_0$, Alfv\'en frequency $\omega_\mathrm{A}$ and Brunt-Brunt-V\"ais\"al\"a
frequency $N$, with the the correct order of $N > \Omega_0 > \omega_\mathrm{A}$
that typically exists in realistic stars.

We find that as theoretically expected, the initial conditions adopted are
unstable to the Taylor instability. The $m=1$ mode clearly dominates the linear
growth of the instability, and the linear growth rates are well-predicted by the
linear eigenvalue calculations. The linear growth phase is later accompanied by
a non-linear coupling between the $m=1$ and other azimuthal modes, leading to
the growth of $m=0$ and $m\geq2$ modes. The $m=0$ component of the poloidal
field is amplified by the instability, signaling a dynamo that can regenerate
the poloidal field as necessary for the Tayler-Spruit dynamo to occur.

Both the linear growth rates and the saturated magnetic field measured in the
simulations scale strongly with angular velocities $\Omega_0$ and the Alfv\'en
frequencies $\omega_\mathrm{A}$ parameters, and are nearly independent of $N$.
While this has been predicted from linear theory and non-linear saturation
models, the scaling in our simulations is steeper, due to the fact they are not
in the asymptotic limit of $\omega_{\rm A} \ll \Omega_0 \ll N$ assumed in
analytic work. With greater scale separations, the linear eigenvalue
calculations are able to replicate the scaling relations of $\gamma \propto
\omega_\mathrm{A}^2 \Omega_0^{-1}$ predicted by analytic work, even though the
resulted growth rates are prohibitively too small to simulate numerically.

The linear growth is ultimately followed by a non-linear saturation of the
instability, which appears to be caused by secondary shear instabilities. The
saturation is also accompanied by an inward migration of unstable motions,
whose cause is not clear. We argued that saturation via secondary shear
instability is unlikely to operate in real stars where the stratification is
greater (preventing shear instabilities), and where Alfv\'en wave damping
becomes more important due to the larger separation of scales than we could
achieve in our simulations. During the saturated phase, energy is dissipated
primarily through magnetic diffusion. However, it is dissipated much more slowly
than predicted by the model of \citet{spruit:02}, likely entailing the
instability can grow to larger amplitudes as expected in the model of
\citet{fuller:19}.

Certain caveats apply to this study. First, due to limited computational
capability, the scale separations between the parameters $\Omega_0$,
$\omega_\mathrm{A}$ and $N$ are much less than that in real stars, the scaling
relations obtained from simulations might differ from what occurs in realistic
stellar environments. Second, although amplification of the axisymmetric
poloidal magnetic field is observed in the simulations, differential rotation is
not included in our simulations, which is necessary to close the loop of the
Tayler-Spruit dynamo. We thus cannot directly test theoretical predictions of
the angular momentum transport caused by this dynamo, leaving room for
improvement in future work.

\section*{Acknowledgements}
The authors thank the referee Florence Marcotte for providing a
constructive report which greatly improves this paper. We also thank Matteo
Cantiello, Adam Jermyn and Eliot Quataert for helpful comments and discussions.
SJ is supported by the Natural Science Foundation of China (grants 12133008,
12192220, and 12192223), the science research grants from the China Manned Space
Project (No. CMS-CSST-2021-B02) and a Sherman Fairchild Fellowship from Caltech.
JF is thankful for support through an Innovator Grant from The Rose Hills
Foundation, and the Sloan Foundation through grant FG-2018-10515. DL is
supported in part by NASA HTMS grant 80NSSC20K1280. The simulations were
performed on the Stampede2 under the XSEDE allocation AST200022, the High
Performance Computing Resource in the Core Facility for Advanced Research
Computing at Shanghai Astronomical Observatory and the Wheeler cluster at
Caltech. This research was supported in part by the National Science Foundation
under Grant No. NSF PHY-1748958. We have made use of NASA's Astrophysics Data
System. Data analysis and visualization are made with {\small Python 3}, and its
packages including {\small NumPy} \citep{van2011numpy}, {\small SciPy}
\citep{oliphant2007python}, {\small Matplotlib} \citep{hunter2007matplotlib} and
the {\small yt} astrophysics analysis software suite \citep{Turk2010}.

\dataavailability{The data supporting the plots within this article are available on reasonable request to the corresponding author.} 

\bibliography{ms_extracted}

\end{document}